\documentclass[prd,onecolumn,showpacs,superscriptaddress,nofootinbib,floatfix,12pt]{revtex4-1} 
\usepackage[utf8]{inputenc}
\usepackage{color}
\usepackage{amsfonts,amsmath,amssymb}
\usepackage{graphicx}
\usepackage{bm}
\usepackage{hyperref}
\hypersetup{colorlinks = true,
           allcolors = blue}
\usepackage{color}
\usepackage{epstopdf}

\newcommand{\sdstarpo}{\Sigma_c^+\bar D^{*0}}
\newcommand{\sdstarppm}{\Sigma_c^{++} D^{*-}}
\newcommand{\sdstar}{\Sigma_c \bar D^{*}}
\newcommand{\pchigh}{P_c(4457)}
\newcommand{\pchighp}{P_c(4457)^+}
\newcommand{\pclow}{P_c(4312)}
\newcommand{\pclowp}{P_c(4312)^+}
\newcommand{\pcmid}{P_c(4440)}
\newcommand{\pcmidp}{P_c(4440)^+}
\newcommand{\br}{\mathcal{B}}
\newcommand{\order}[1]{\mathcal{O}(#1)}
\newcommand{\jp}{J/\psi p}
\newcommand{\jdelta}{J/\psi \Delta^+}
\newcommand{\mev}{\text{MeV}}
\newcommand{\amp}{\mathcal{A}}

\newcommand{\rdeltap}{R_{\Delta^+/p}}

\newcommand{\itp}{\affiliation{CAS Key Laboratory of Theoretical Physics, Institute of Theoretical Physics, Chinese Academy of Sciences, Beijing 100190, China}}

\newcommand{\ucas}{\affiliation{School of Physical Sciences, University of Chinese Academy of Sciences, Beijing 100049, China}}

\newcommand{\ubonn}{\affiliation{Helmholtz-Institut f\"ur Strahlen- und              Kernphysik and Bethe Center for Theoretical Physics,\\ 
Universit\"at Bonn,  D-53115 Bonn, Germany}}
             
\newcommand{\fzj}{\affiliation{Institute for Advanced Simulation, Institut f\"ur Kernphysik and J\"ulich Center for Hadron Physics,\\
Forschungszentrum J\"ulich, D-52425 J\"ulich, Germany}}

\newcommand{\tsu}{\affiliation{Tbilisi State  University,  0186 Tbilisi, Georgia}}

\begin{document}

\title{Isospin breaking decays as a diagnosis of the hadronic molecular structure of the \boldmath{$\pchigh$}}

\author{Feng-Kun Guo}\email[E-mail address: ]{fkguo@itp.ac.cn}
\itp\ucas

\author{Hao-Jie Jing}\email[E-mail address: ]{jinghaojie@itp.ac.cn}\itp\ucas

\author{Ulf-G. Mei{\ss}ner}\email[E-mail address: ]{meissner@hiskp.uni-bonn.de}\ubonn\fzj\tsu

\author{Shuntaro Sakai}\email[E-mail address: ]{shsakai@itp.ac.cn}\itp

\begin{abstract}
The LHCb Collaboration announced the observation of three narrow structures consistent with hidden-charm pentaquark states. They are candidates of hadronic molecules formed of a pair of a charmed baryon and an anticharmed meson. Among them, the $P_c(4457)$ mass is consistent with earlier predictions of a $\Sigma_c\bar D^*$ molecule with $I=1/2$. We point out that if such a picture were true, one would have $\mathcal{B}(P_c(4457)\to J/\psi \Delta^+)/\mathcal{B}(P_c(4457)\to J/\psi p)$ at the level ranging from a few percent to about 30\%. Such a large isospin breaking decay ratio is two to three orders of magnitude larger than that for normal hadron resonances. It is a unique feature of the $\sdstar$ molecular model, and can be checked by LHCb.
\end{abstract}

\date{\today}

\maketitle

\vspace{2cm}


Four years after the discovery of the hidden-charm pentaquark-like states $P_c(4380)$ and $P_c(4450)$~\cite{Aaij:2015tga}, with the full set of Run-1 and Run-2 data the LHCb Collaboration announced  the observation of more structures consistent with hidden-charm pentaquark states with masses and widths given by~\cite{LHCb:2019,Aaij:2019vzc}
\begin{align}
 M_{\pclowp} &= 4311.9\pm0.7^{+6.8}_{-0.6}~\mev, & \Gamma_{\pclowp} &= 9.8\pm2.7^{+3.7}_{-4.5}~\mev, \nonumber\\
 M_{\pcmidp} &= 4440.3\pm1.3^{+4.1}_{-4.7}~\mev, & \Gamma_{\pcmidp} &= 20.6\pm4.9^{+8.7}_{-10.1}~\mev,\nonumber\\
 M_{\pchighp} &= 4457.3\pm0.6^{+4.1}_{-1.7}~\mev, & \Gamma_{\pchighp} &= 6.4\pm2.0^{+5.7}_{-1.9}~\mev.
\end{align}
That is, the $P_c(4450)$ reported earlier is split into two peaks corresponding to the $\pcmid$ and the $\pchigh$, and the small spike (sticking out in a single bin) at slightly above 4.3~GeV in the 2015 measurement is resolved into a pronounced peak with a 7.3$\sigma$ significance. Pentaquark states with hidden-charm as hadronic molecules of a pair of a charmed baryon and an anticharmed meson were predicted to exist in this mass region prior to the LHCb observations~\cite{Wu:2010jy,Wang:2011rga,Yang:2011wz,Yuan:2012wz,Wu:2012md,Xiao:2013yca,Uchino:2015uha,Karliner:2015ina}. In particular, the masses of the $\pclow$ and $\pchigh$ are in a remarkable agreement with the predictions for the isospin $I=1/2$ $\Sigma_c\bar D$ ($J^P=1/2^-$) and $\Sigma_c\bar D^*$ ($J^P=1/2^-$ or $3/2^-$) $S$-wave bound states in Ref.~\cite{Wu:2012md} where a coupled-channel formalism with the vector-meson exchange potential is used. 
The first observation in Ref.~\cite{Aaij:2015tga} inspired a flood of models for the $P_c$ structures, such as the baryon--meson
molecules~\cite{Chen:2015loa,Chen:2015moa,Roca:2015dva,He:2015cea,Meissner:2015mza,Burns:2015dwa, Xiao:2015fia,Lu:2016nnt,Shen:2016tzq,Yamaguchi:2016ote,Azizi:2016dhy,Lin:2017mtz,Geng:2017hxc,Yamaguchi:2017zmn,Wang:2018waa,Liu:2018zzu}, compact pentaquark 
states~\cite{Maiani:2015vwa,Lebed:2015tna,Li:2015gta,Mironov:2015ica,
Anisovich:2015cia,Wang:2015epa,Ali:2016dkf,Takeuchi:2016ejt}  and
baryocharmonia~\cite{Kubarovsky:2015aaa}, while the importance of triangle singularities, in particularly for the $P_c(4450)$, has
also been discussed~\cite{Guo:2015umn,Liu:2015fea,Guo:2016bkl,Bayar:2016ftu}.\footnote{It could be that the triangle singularities enhance the production of the $P_c$ states at around 4.45~GeV.} Reviews of these models can be found in Refs.~\cite{Chen:2016qju,Lebed:2016hpi,Esposito:2016noz,Guo:2017jvc,Ali:2017jda,Olsen:2017bmm,Karliner:2017qhf,Cerri:2018ypt}. Of particular interest here is the interpretation of the $P_c(4450)$ as an $I=1/2$ $\sdstar$ molecular with  $J^P=3/2^-$~\cite{Roca:2015dva,He:2015cea,Xiao:2015fia,Burns:2015dwa,Lu:2016nnt,Lin:2017mtz,Geng:2017hxc,Wang:2018waa,Liu:2018zzu} (see also early predictions in Ref.~\cite{Wu:2012md}), which is adopted as the interpretation for the $\pchighp$ in Refs.~\cite{Chen:2019bip,Chen:2019asm}. We notice that such an interpretation will lead to large isospin breaking effects in the decays. We have the following nearby $\Sigma_c\bar D^*$ thresholds:\footnote{As noticed in Ref.~\cite{LHCb:2019}, the mass of the $\pchighp$ coincides with the $\Lambda_c(2595)^+\bar D^0$ threshold, $4457.09\pm0.28$~MeV. }
\begin{equation}
    M_{\Sigma_c^{+}} + M_{\bar D^{*0}} = 4459.8\pm0.4~\mev, \qquad 
    M_{\Sigma_c^{++}} + M_{D^{*-}} = 4464.23\pm0.15~\mev.
\end{equation}
Thus, the binding energy of the $\pchighp$ with respect to the $\sdstarpo$ threshold, $2.5^{+1.8}_{-4.2}$~MeV, is sizably smaller than that with respect to the $\sdstarppm$ threshold, $6.9^{+1.8}_{-4.1}$~MeV. As a result, one would expect sizeable isospin breaking effects in the decays, similar to the case of the $X(3872)$ which decays with comparable rates into the $I=0$ $J/\psi \pi^+\pi^-\pi^0$ and $I=1$ $J/\psi \pi^+\pi^-$  final states,\footnote{Several interesting similarities between the $P_c(4450)$ and the $X(3872)$, including the possibility of a sizeable isospin symmetry breaking, were discussed in Ref.~\cite{Burns:2015dwa}.} though with a much more modest magnitude as will be shown in the following.

\begin{figure}[tb]
    \centering
    \includegraphics[width=0.95\textwidth]{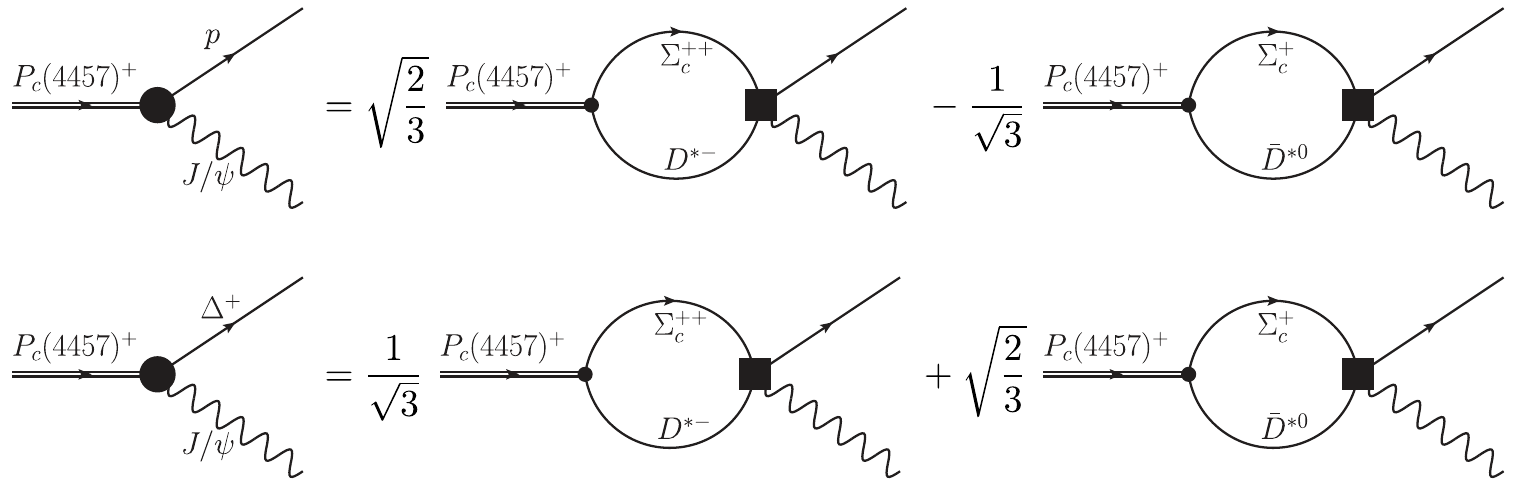}
    \caption{Illustration of the decays of the $\pchighp\to \jp$ and $\pchighp\to \jdelta$ through $\sdstar$ loops. Here the double lines represent the physical $P_c(4457)^+$ state.}
    \label{fig:decays}
\end{figure}

Since the isospin of the $\Sigma_c$ is 1 and that of the $\bar D^*$ is $1/2$, one can form $I=3/2$ and $I=1/2$ states out of them,
\begin{align}
    \left| \sdstar; I=\frac12,I_3=\frac12 \right\rangle &= \sqrt{\frac23} \left| \sdstarppm \right\rangle - \frac{1}{\sqrt{3}} \left| \sdstarpo \right\rangle, \nonumber\\
    \left| \sdstar; I=\frac32,I_3=\frac12 \right\rangle &= \frac{1}{\sqrt{3}} \left| \sdstarppm \right\rangle + \sqrt{\frac23} \left| \sdstarpo \right\rangle.
    \label{eq:isospin}
\end{align}
In the $\sdstar$ molecular picture, the decays of the $\pchighp$ into the $\jp$ and $\jdelta$ dominantly proceed through the $\sdstar$ loops with the intermediate states carrying different electric charges, as shown in Fig.~\ref{fig:decays}.
We denote the $S$-wave coupling constant for the $\pchighp \to\sdstarpo$ vertex as $g_{+,0}$ and that for the  $\pchighp \to\sdstarppm$ vertex as $g_{++,-}$. Assuming the $\pchigh$ to be a hadronic molecule generated from the $I=1/2$ $S$-wave interaction between the $\sdstar$ pair, 
from the Lippmann-Schwinger equation (LSE), we have 
\begin{equation}
    T^{-1}_{1/2} = V_{1/2}^{-1}-  G_{\sdstar}^\Lambda = 0
    \label{eq:lse}
\end{equation}
when the energy equals to the mass of the $P_c$. Here, $T_{1/2}$ is the $I=1/2$ $\sdstar$ scattering $T$-matrix, $V_{1/2}$ is the corresponding nonrelativistic potential, and $G_{\sdstar}^\Lambda$ is the $\sdstar$ two-body Green's function whose form is irrelevant here (it will be given below when it is used). In $G_{\sdstar}^\Lambda$, the isospin averaged masses for the $\Sigma_c$ and $\bar{D}^*$ should be used. Now let us switch on isospin breaking and consider the two-channel ($\sdstarppm, \sdstarpo$) nonrelativistic system. Because the products of couplings are the residues of the $T$-matrix elements, i.e., $g_{++,-}^2=\operatorname{Res} T_{++,-\to ++,-}$ and $g_{++,-} g_{+,0}=\operatorname{Res} T_{++,-\to +,0}$, we get from the two-channel LSE the following ratio (the energy is at the $P_c$ mass),
\begin{align}
    \frac{g_{++,-}}{g_{+,0}} = \frac{2V_{1/2} + V_{3/2} - 3 V_{1/2} V_{3/2} G_{+,0}^\Lambda}{-\sqrt{2}\left(V_{1/2}-V_{3/2}\right)},
    \label{eq:gratio0}
\end{align}
where $V_{3/2}$ is the potential for the $I=3/2$ $\sdstar$ scattering, $G_{+,0}^\Lambda$ is the two-body Green's function for $\sdstarpo$, and we have used Eq.~\eqref{eq:isospin} to express the particle-basis potentials in terms of the isospin-basis ones $V_{1/2}$ and $V_{3/2}$. From Eq.~\eqref{eq:lse}, we get $V_{1/2}G_{+,0}^\Lambda = 1 - V_{1/2} \left(G_{+,0}^\Lambda - G_{\sdstar}^\Lambda\right)$, where the second term is an isospin breaking effect and is much smaller than 1. Therefore, Eq.~\eqref{eq:gratio0} becomes
\begin{equation}
    g_{++,-} \simeq -\sqrt{2} g_{+,0} \,.
    \label{eq:couplings}
\end{equation}
Then from Fig.~\ref{fig:decays} one sees that in the isospin limit when all the masses in the same isospin multiplet are degenerate, the two loops exactly cancel with each other for the decay into the $I=3/2$ final state $\jdelta$. The isospin splittings of the intermediate particles make the transition possible. In order to estimate the size of the isospin breaking effect, we make use of the method of Ref.~\cite{Gamermann:2009uq} which was developed for the $X(3872)$ (see also Refs.~\cite{Hanhart:2011tn,Li:2012cs}).

The magnitudes of the three-momenta for the decays of the $\pchighp$ into $\jp$ and $\jdelta$ are about 0.83~GeV and 0.52~GeV, respectively. They are much larger than the binding momenta which are $73$~MeV and 124~MeV for the $\sdstarpo$ and $\sdstarppm$, respectively (here the central values of all involved masses are used).
Thus, these decays are short-distance processes, and the decay rates would be determined by the wave function at the origin.

The wave function at the origin for a two-body component (labeled by $i$) of a physical state with a mass $M$  is given by
\begin{equation}
    \psi_i(r= 0) =  \int \frac{d^3\vec q}{(2\pi)^3} \langle \vec q\,|\psi\rangle = -2\mu_i \int\frac{d^3\vec q}{(2\pi)^3} \frac{ \langle \vec q\,|\hat V_i| \psi\rangle}{ \gamma_i^2 + \vec q^{\,2} },
\end{equation}
where we have used the Schr\"odinger equation $\left(\vec q^{\,2}/(2\mu_i) + \hat V_i \right)|\psi\rangle = (M-m_1-m_2)|\psi\rangle$, and the binding momentum is defined as $\gamma_i = \sqrt{2\mu_i(m_1+m_2-M)}$,  with $m_{1,2}$ the masses of the constituents and $\mu_i=m_1 m_2/(m_1+m_2)$ the reduced mass. Since the physical state is nearby the threshold, one can approximate the $S$-wave vertex form factor $\langle \vec q\,|\hat V_i| \psi\rangle$ by the coupling constant $g_i$. Then, one gets
\begin{equation}
    \psi^{\Lambda}_{i}(r= 0) = -2\mu_i g_i \int\frac{d^3\vec q}{(2\pi)^3} \frac{ \exp(-2 \vec q^{\,2}/\Lambda^2)}{ \gamma_i^2 + \vec q^{\,2} } \equiv g_i G^{\Lambda}_{i},
\end{equation}
where a Gaussian form factor with a cutoff $\Lambda$ is introduced to regularize the ultraviolet  divergence, and $G^{\Lambda}_{i}$ is simply the nonrelativistic two-point scalar loop integral evaluated at the mass of the state. When $M<m_1+m_2$, it is given by 
\begin{equation}
    G^\Lambda_i = - \frac{\mu_i \Lambda}{(2\pi)^{3/2}} - \frac{\mu_i\gamma_i}{2\pi} e^{2\gamma_i^2 /\Lambda^2} \left[ \text{erf}\left(\frac{\sqrt{2}\gamma_i }{\Lambda}\right) -1 \right],
\end{equation}
where $\text{erf}\,(x)$ is the error function.

Thus, for the $\pchighp$ we have
\begin{equation}
    \psi^\Lambda_{++,-}(r=0) = g_{++,-}G^\Lambda_{++,-}\,, \quad \text{and} \quad
    \psi^\Lambda_{+,0}(r=0) = g_{+,0}G^\Lambda_{+,0}\,,
    \label{eq:wavfun}
\end{equation}
for the $\sdstarppm$ and $\sdstarpo$ components, respectively. 
From Eq.~\eqref{eq:isospin}, the isospin $I=1/2$ and $I=3/2$ components are 
\begin{align}
    \psi^\Lambda_{1/2} (r=0) & = \sqrt{\frac23}  \psi^\Lambda_{++,-}(r=0) - \frac1{\sqrt{3}} \psi^\Lambda_{+,0}(r=0), \nonumber\\
    \psi^\Lambda_{3/2} (r=0) & = \frac1{\sqrt{3}} \psi^\Lambda_{++,-}(r=0) + \sqrt{\frac23}  \psi^\Lambda_{+,0}(r=0) .
\end{align}

\begin{figure}[tb]
    \centering
    \includegraphics[width=0.65\textwidth]{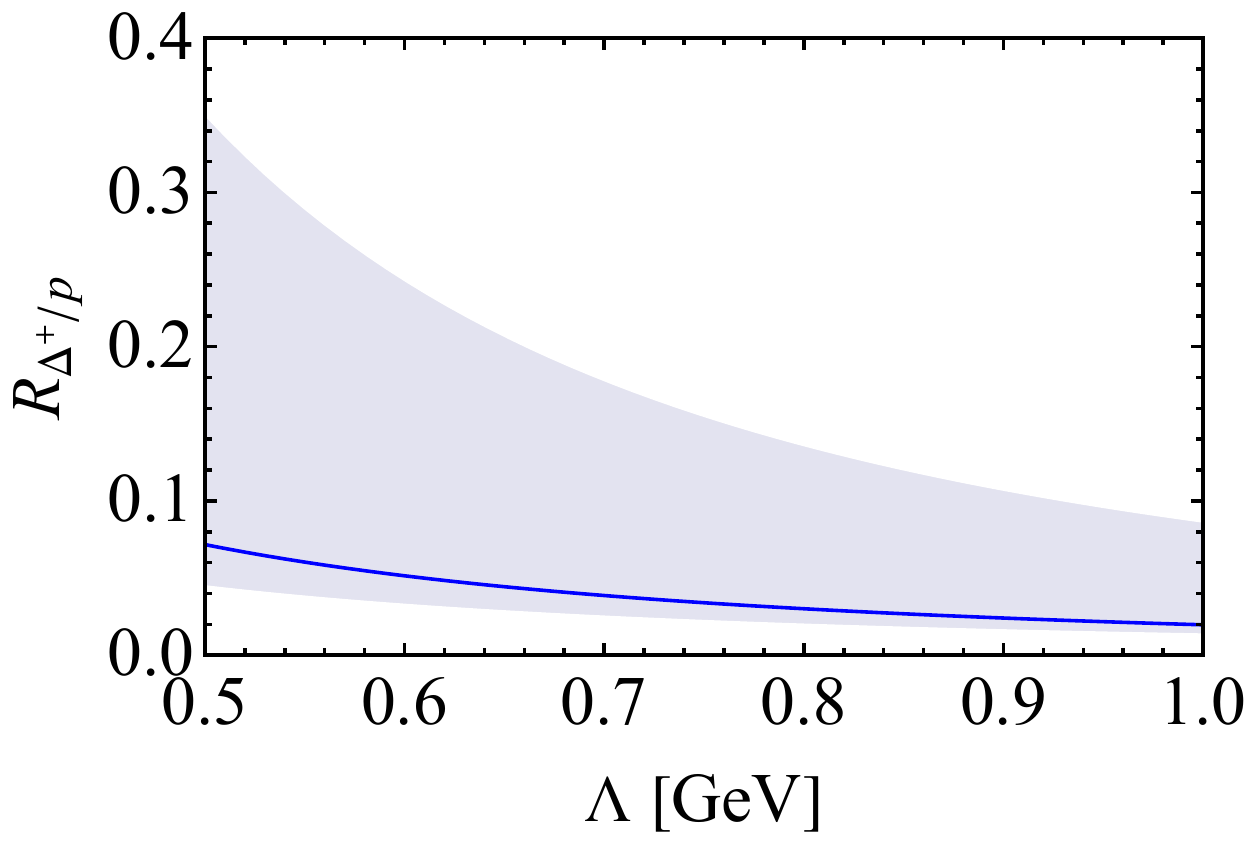}
    \caption{Dependence of the ratio $\rdeltap$ on the cutoff. The solid line corresponds to the result calculated using the central values of all the involved masses, and the band reflects the uncertainties of the masses.}
    \label{fig:ratio}
\end{figure}

In view that the $\Delta$ resonances and the nucleons are in the same spin-flavor multiplet in the large $N_c$ limit (see, e.g., Ref.~\cite{Dashen:1993jt}), one gets the following relation for the decay amplitudes
\begin{equation}
    \frac{ \left|\amp(\pchighp\to \jdelta)\right|}{ \left|\amp(\pchighp\to \jp)\right|} \simeq 
    \sqrt{10}\left| \frac{\psi^\Lambda_{3/2}(r=0) }{ \psi^\Lambda_{1/2}(r=0)} \right| = 2\sqrt{5} \left | \frac{G^\Lambda_{++,-} - G^\Lambda_{+,0}}{ 2G^\Lambda_{++,-} + G^\Lambda_{+,0}} \right|,
\end{equation} 
where the factor of $\sqrt{10}$ comes from the spin-flavor matrix elements worked out in Ref.~\cite{Burns:2015dwa} (see Eqs.~(17,18) therein), and Eq.~\eqref{eq:couplings} has been used. We have further assumed that $V_{3/2}$ is much smaller than $V_{1/2}$ so that we can neglect contribution from the isospin breaking effect in Eq.~\eqref{eq:couplings} here. This is plausible in the molecular picture as the $I=1/2$ interaction needs to be strong to produce the $P_c(4457)$ as a $\sdstar$ bound state.\footnote{For the $X(3872)$ in the $D\bar D^*+c.c.$ hadronic molecular picture, the $I=1$ potential is indeed much weaker than the $I=0$ one, see, e.g., Ref.~\cite{Guo:2014hqa}.} From this equation,
and taking into account the $S$-wave phase spaces for the decays of the $\pchighp$ into the $\jp$ and $\jdelta$, 
one can predict the isospin breaking ratio
\begin{equation}
    R_{\Delta^+/p} \equiv \frac{\br(\pchighp\to \jdelta)}{ \br(\pchighp\to \jp)},
\end{equation}
and the result is shown in Fig.~\ref{fig:ratio} with the cutoff $\Lambda$ in the region from 0.5~GeV to 1~GeV. 
One sees that the ratio ranges from a few percent to as large as 30\%  with the large uncertainty mainly from the uncertainty of the $\pchighp$ mass. It is two to three orders of magnitude larger than the isospin breaking effects for the decays of normal hadron resonances. In order to see that, one notices that there are two sources of isospin breaking: the up and down quark mass difference, and the electromagnetic interactions (virtual photons). 
They give amplitudes of the order of $(m_d-m_u)/\Lambda_\text{QCD}$ and $\alpha$, respectively, where $\Lambda_\text{QCD}$ is the nonperturbative scale in quantum chromodynamics and $\alpha$ is the fine structure constant. Both of them are of $\order{10^{-2}}$, and thus lead to a suppression for the branching fractions of $\order{10^{-4}}$. To give an example, the ratio of the branching fraction of the decay of an isoscalar state into another isoscalar and a $\pi^0$ over that into the same isoscalar and an $\eta$ is given by $\epsilon_{\pi^0\eta}^2$ up to the phase space factor. The isospin breaking $\pi^0$-$\eta$ mixing angle is 
\begin{equation}
    \epsilon_{\pi^0\eta} = \frac{\sqrt{3}}{2} \frac{m_d-m_u}{2m_s - m_u-m_d} \simeq \frac{\sqrt{3}}{2} \frac{ M_{K^0}^2 - M_{K^\pm}^2 - M_{\pi^0}^2 + M_{\pi^\pm}^2 }{ M_{K^0}^2 + M_{K^\pm}^2 - M_{\pi^0}^2 - M_{\pi^\pm}^2} \simeq 0.01 \,,
\end{equation}
where the combinations of meson masses are constructed such that the virtual photon effects are canceled out. 

To summarize, in this paper we propose that the structure of the $\pchigh$ can be diagnosed using isospin breaking decays. If the $\pchighp$ is an $S$-wave $\sdstar$ hadronic molecule with $I=1/2$, which implies that it couples most strongly to the $\sdstar$ channels, then because its mass is closer to the $\sdstarpo$ threshold than to the $\sdstarppm$ one, one expects large isospin breaking effects in its decays. 
A quantitative estimate of the ratio $\mathcal{B}(\pchighp\to \jdelta)/\mathcal{B}(\pchighp\to \jp)$ gives a value ranging from $\order{10^{-2}}$ to about 30\%, where the large uncertainty comes mainly from the mass of the $\pchighp$. It is two to three orders of magnitude higher than the isospin breaking effects for the decays of normal hadron resonances. 
It is worthwhile to mention that the large isospin breaking effect is a key to unveiling the nature   of the $D_{s0}^*(2317)$, whose isospin breaking decay width is about $100~\text{keV}$~\cite{Faessler:2007gv,Lutz:2007sk,Guo:2008gp,Liu:2012zya,Guo:2018kno} in the $DK$ molecular picture and is one order of magnitude smaller~\cite{Bardeen:2003kt,Colangelo:2003vg} if it couples weakly to the $DK$ (for detailed discussions, see Ref.~\cite{Guo:2017jvc}). 
Therefore, we suggest to search for the $\pchighp$ ($\pchigh^0$) in the $\jdelta (J/\psi \Delta^0)$ mode. Given the large ratio, it is feasible at the LHCb experiment.

\medskip

\begin{acknowledgements}
F.-K. G. is grateful for the hospitality of the Helmholtz-Institut f\"ur Strahlen- und       Kernphysik (HISKP) where this work was done.
This work is supported in part by the National Natural Science Foundation of China (NSFC) and  the Deutsche Forschungsgemeinschaft (DFG) through the funds provided to the Sino-German Collaborative Research Center ``Symmetries and the Emergence of Structure in QCD"  (NSFC Grant No. 11621131001, DFG Grant No. TRR110), by the NSFC under Grant No. 11747601 and No. 11835015, by the Chinese Academy of Sciences (CAS) under Grant No. QYZDB-SSW-SYS013 and No. XDPB09, by
the CAS Center for Excellence in Particle Physics (CCEPP),  by the CAS President’s International Fellowship Initiative (PIFI) (Grant No.~2018DM0034 and No.~2019PM0108), and by the VolkswagenStiftung (Grant No. 93562).
\end{acknowledgements}

\medskip

\bibliography{Pc}

\begin{thebibliography}{62}%
\makeatletter
\providecommand \@ifxundefined [1]{%
 \@ifx{#1\undefined}
}%
\providecommand \@ifnum [1]{%
 \ifnum #1\expandafter \@firstoftwo
 \else \expandafter \@secondoftwo
 \fi
}%
\providecommand \@ifx [1]{%
 \ifx #1\expandafter \@firstoftwo
 \else \expandafter \@secondoftwo
 \fi
}%
\providecommand \natexlab [1]{#1}%
\providecommand \enquote  [1]{``#1''}%
\providecommand \bibnamefont  [1]{#1}%
\providecommand \bibfnamefont [1]{#1}%
\providecommand \citenamefont [1]{#1}%
\providecommand \href@noop [0]{\@secondoftwo}%
\providecommand \href [0]{\begingroup \@sanitize@url \@href}%
\providecommand \@href[1]{\@@startlink{#1}\@@href}%
\providecommand \@@href[1]{\endgroup#1\@@endlink}%
\providecommand \@sanitize@url [0]{\catcode `\\12\catcode `\$12\catcode
  `\&12\catcode `\#12\catcode `\^12\catcode `\_12\catcode `\%12\relax}%
\providecommand \@@startlink[1]{}%
\providecommand \@@endlink[0]{}%
\providecommand \url  [0]{\begingroup\@sanitize@url \@url }%
\providecommand \@url [1]{\endgroup\@href {#1}{\urlprefix }}%
\providecommand \urlprefix  [0]{URL }%
\providecommand \Eprint [0]{\href }%
\providecommand \doibase [0]{http://dx.doi.org/}%
\providecommand \selectlanguage [0]{\@gobble}%
\providecommand \bibinfo  [0]{\@secondoftwo}%
\providecommand \bibfield  [0]{\@secondoftwo}%
\providecommand \translation [1]{[#1]}%
\providecommand \BibitemOpen [0]{}%
\providecommand \bibitemStop [0]{}%
\providecommand \bibitemNoStop [0]{.\EOS\space}%
\providecommand \EOS [0]{\spacefactor3000\relax}%
\providecommand \BibitemShut  [1]{\csname bibitem#1\endcsname}%
\let\auto@bib@innerbib\@empty
\bibitem [{\citenamefont {Aaij}\ \emph {et~al.}(2015)\citenamefont {Aaij} \emph
  {et~al.}}]{Aaij:2015tga}%
  \BibitemOpen
  \bibfield  {author} {\bibinfo {author} {\bibfnamefont {R.}~\bibnamefont
  {Aaij}} \emph {et~al.} (\bibinfo {collaboration} {LHCb}),\ }\href {\doibase
  10.1103/PhysRevLett.115.072001} {\bibfield  {journal} {\bibinfo  {journal}
  {Phys. Rev. Lett.}\ }\textbf {\bibinfo {volume} {115}},\ \bibinfo {pages}
  {072001} (\bibinfo {year} {2015})},\ \Eprint
  {http://arxiv.org/abs/1507.03414} {arXiv:1507.03414 [hep-ex]} \BibitemShut
  {NoStop}%
\bibitem [{\citenamefont {Skwarnicki}()}]{LHCb:2019}%
  \BibitemOpen
  \bibfield  {author} {\bibinfo {author} {\bibfnamefont {T.}~\bibnamefont
  {Skwarnicki}},\ }\href@noop {} {\enquote {\bibinfo {title} {{Hadron
  spectroscopy and exotic states at LHCb}},}\ }\bibinfo {howpublished}
  {\url{http://moriond.in2p3.fr/QCD/2019/TuesdayMorning/Skwarnicki.pptx}},\
  \bibinfo {note} {talk given on behalf of the LHCb Collaboration at the QCD
  and High Energy Interactions at the 54th Rencontres de Moriond, 23-30 March
  2019}\BibitemShut {NoStop}%
\bibitem [{\citenamefont {Aaij}\ \emph {et~al.}(2019)\citenamefont {Aaij} \emph
  {et~al.}}]{Aaij:2019vzc}%
  \BibitemOpen
  \bibfield  {author} {\bibinfo {author} {\bibfnamefont {R.}~\bibnamefont
  {Aaij}} \emph {et~al.} (\bibinfo {collaboration} {LHCb}),\ }\href@noop {} {\
  (\bibinfo {year} {2019})},\ \Eprint {http://arxiv.org/abs/1904.03947}
  {arXiv:1904.03947 [hep-ex]} \BibitemShut {NoStop}%
\bibitem [{\citenamefont {Wu}\ \emph {et~al.}(2010)\citenamefont {Wu},
  \citenamefont {Molina}, \citenamefont {Oset},\ and\ \citenamefont
  {Zou}}]{Wu:2010jy}%
  \BibitemOpen
  \bibfield  {author} {\bibinfo {author} {\bibfnamefont {J.-J.}\ \bibnamefont
  {Wu}}, \bibinfo {author} {\bibfnamefont {R.}~\bibnamefont {Molina}}, \bibinfo
  {author} {\bibfnamefont {E.}~\bibnamefont {Oset}}, \ and\ \bibinfo {author}
  {\bibfnamefont {B.~S.}\ \bibnamefont {Zou}},\ }\href {\doibase
  10.1103/PhysRevLett.105.232001} {\bibfield  {journal} {\bibinfo  {journal}
  {Phys. Rev. Lett.}\ }\textbf {\bibinfo {volume} {105}},\ \bibinfo {pages}
  {232001} (\bibinfo {year} {2010})},\ \Eprint {http://arxiv.org/abs/1007.0573}
  {arXiv:1007.0573 [nucl-th]} \BibitemShut {NoStop}%
\bibitem [{\citenamefont {Wang}\ \emph {et~al.}(2011)\citenamefont {Wang},
  \citenamefont {Huang}, \citenamefont {Zhang},\ and\ \citenamefont
  {Zou}}]{Wang:2011rga}%
  \BibitemOpen
  \bibfield  {author} {\bibinfo {author} {\bibfnamefont {W.~L.}\ \bibnamefont
  {Wang}}, \bibinfo {author} {\bibfnamefont {F.}~\bibnamefont {Huang}},
  \bibinfo {author} {\bibfnamefont {Z.~Y.}\ \bibnamefont {Zhang}}, \ and\
  \bibinfo {author} {\bibfnamefont {B.~S.}\ \bibnamefont {Zou}},\ }\href
  {\doibase 10.1103/PhysRevC.84.015203} {\bibfield  {journal} {\bibinfo
  {journal} {Phys. Rev.}\ }\textbf {\bibinfo {volume} {C84}},\ \bibinfo {pages}
  {015203} (\bibinfo {year} {2011})},\ \Eprint {http://arxiv.org/abs/1101.0453}
  {arXiv:1101.0453 [nucl-th]} \BibitemShut {NoStop}%
\bibitem [{\citenamefont {Yang}\ \emph {et~al.}(2012)\citenamefont {Yang},
  \citenamefont {Sun}, \citenamefont {He}, \citenamefont {Liu},\ and\
  \citenamefont {Zhu}}]{Yang:2011wz}%
  \BibitemOpen
  \bibfield  {author} {\bibinfo {author} {\bibfnamefont {Z.-C.}\ \bibnamefont
  {Yang}}, \bibinfo {author} {\bibfnamefont {Z.-F.}\ \bibnamefont {Sun}},
  \bibinfo {author} {\bibfnamefont {J.}~\bibnamefont {He}}, \bibinfo {author}
  {\bibfnamefont {X.}~\bibnamefont {Liu}}, \ and\ \bibinfo {author}
  {\bibfnamefont {S.-L.}\ \bibnamefont {Zhu}},\ }\href {\doibase
  10.1088/1674-1137/36/1/002, 10.1088/1674-1137/36/3/006} {\bibfield  {journal}
  {\bibinfo  {journal} {Chin. Phys.}\ }\textbf {\bibinfo {volume} {C36}},\
  \bibinfo {pages} {6} (\bibinfo {year} {2012})},\ \Eprint
  {http://arxiv.org/abs/1105.2901} {arXiv:1105.2901 [hep-ph]} \BibitemShut
  {NoStop}%
\bibitem [{\citenamefont {Yuan}\ \emph {et~al.}(2012)\citenamefont {Yuan},
  \citenamefont {Wei}, \citenamefont {He}, \citenamefont {Xu},\ and\
  \citenamefont {Zou}}]{Yuan:2012wz}%
  \BibitemOpen
  \bibfield  {author} {\bibinfo {author} {\bibfnamefont {S.~G.}\ \bibnamefont
  {Yuan}}, \bibinfo {author} {\bibfnamefont {K.~W.}\ \bibnamefont {Wei}},
  \bibinfo {author} {\bibfnamefont {J.}~\bibnamefont {He}}, \bibinfo {author}
  {\bibfnamefont {H.~S.}\ \bibnamefont {Xu}}, \ and\ \bibinfo {author}
  {\bibfnamefont {B.~S.}\ \bibnamefont {Zou}},\ }\href {\doibase
  10.1140/epja/i2012-12061-2} {\bibfield  {journal} {\bibinfo  {journal} {Eur.
  Phys. J.}\ }\textbf {\bibinfo {volume} {A48}},\ \bibinfo {pages} {61}
  (\bibinfo {year} {2012})},\ \Eprint {http://arxiv.org/abs/1201.0807}
  {arXiv:1201.0807 [nucl-th]} \BibitemShut {NoStop}%
\bibitem [{\citenamefont {Wu}\ \emph {et~al.}(2012)\citenamefont {Wu},
  \citenamefont {Lee},\ and\ \citenamefont {Zou}}]{Wu:2012md}%
  \BibitemOpen
  \bibfield  {author} {\bibinfo {author} {\bibfnamefont {J.-J.}\ \bibnamefont
  {Wu}}, \bibinfo {author} {\bibfnamefont {T.~S.~H.}\ \bibnamefont {Lee}}, \
  and\ \bibinfo {author} {\bibfnamefont {B.~S.}\ \bibnamefont {Zou}},\ }\href
  {\doibase 10.1103/PhysRevC.85.044002} {\bibfield  {journal} {\bibinfo
  {journal} {Phys. Rev.}\ }\textbf {\bibinfo {volume} {C85}},\ \bibinfo {pages}
  {044002} (\bibinfo {year} {2012})},\ \Eprint {http://arxiv.org/abs/1202.1036}
  {arXiv:1202.1036 [nucl-th]} \BibitemShut {NoStop}%
\bibitem [{\citenamefont {Xiao}\ \emph {et~al.}(2013)\citenamefont {Xiao},
  \citenamefont {Nieves},\ and\ \citenamefont {Oset}}]{Xiao:2013yca}%
  \BibitemOpen
  \bibfield  {author} {\bibinfo {author} {\bibfnamefont {C.~W.}\ \bibnamefont
  {Xiao}}, \bibinfo {author} {\bibfnamefont {J.}~\bibnamefont {Nieves}}, \ and\
  \bibinfo {author} {\bibfnamefont {E.}~\bibnamefont {Oset}},\ }\href {\doibase
  10.1103/PhysRevD.88.056012} {\bibfield  {journal} {\bibinfo  {journal} {Phys.
  Rev.}\ }\textbf {\bibinfo {volume} {D88}},\ \bibinfo {pages} {056012}
  (\bibinfo {year} {2013})},\ \Eprint {http://arxiv.org/abs/1304.5368}
  {arXiv:1304.5368 [hep-ph]} \BibitemShut {NoStop}%
\bibitem [{\citenamefont {Uchino}\ \emph {et~al.}(2016)\citenamefont {Uchino},
  \citenamefont {Liang},\ and\ \citenamefont {Oset}}]{Uchino:2015uha}%
  \BibitemOpen
  \bibfield  {author} {\bibinfo {author} {\bibfnamefont {T.}~\bibnamefont
  {Uchino}}, \bibinfo {author} {\bibfnamefont {W.-H.}\ \bibnamefont {Liang}}, \
  and\ \bibinfo {author} {\bibfnamefont {E.}~\bibnamefont {Oset}},\ }\href
  {\doibase 10.1140/epja/i2016-16043-0} {\bibfield  {journal} {\bibinfo
  {journal} {Eur. Phys. J.}\ }\textbf {\bibinfo {volume} {A52}},\ \bibinfo
  {pages} {43} (\bibinfo {year} {2016})},\ \Eprint
  {http://arxiv.org/abs/1504.05726} {arXiv:1504.05726 [hep-ph]} \BibitemShut
  {NoStop}%
\bibitem [{\citenamefont {Karliner}\ and\ \citenamefont
  {Rosner}(2015)}]{Karliner:2015ina}%
  \BibitemOpen
  \bibfield  {author} {\bibinfo {author} {\bibfnamefont {M.}~\bibnamefont
  {Karliner}}\ and\ \bibinfo {author} {\bibfnamefont {J.~L.}\ \bibnamefont
  {Rosner}},\ }\href {\doibase 10.1103/PhysRevLett.115.122001} {\bibfield
  {journal} {\bibinfo  {journal} {Phys. Rev. Lett.}\ }\textbf {\bibinfo
  {volume} {115}},\ \bibinfo {pages} {122001} (\bibinfo {year} {2015})},\
  \Eprint {http://arxiv.org/abs/1506.06386} {arXiv:1506.06386 [hep-ph]}
  \BibitemShut {NoStop}%
\bibitem [{\citenamefont {Chen}\ \emph
  {et~al.}(2015{\natexlab{a}})\citenamefont {Chen}, \citenamefont {Liu},
  \citenamefont {Li},\ and\ \citenamefont {Zhu}}]{Chen:2015loa}%
  \BibitemOpen
  \bibfield  {author} {\bibinfo {author} {\bibfnamefont {R.}~\bibnamefont
  {Chen}}, \bibinfo {author} {\bibfnamefont {X.}~\bibnamefont {Liu}}, \bibinfo
  {author} {\bibfnamefont {X.-Q.}\ \bibnamefont {Li}}, \ and\ \bibinfo {author}
  {\bibfnamefont {S.-L.}\ \bibnamefont {Zhu}},\ }\href {\doibase
  10.1103/PhysRevLett.115.132002} {\bibfield  {journal} {\bibinfo  {journal}
  {Phys. Rev. Lett.}\ }\textbf {\bibinfo {volume} {115}},\ \bibinfo {pages}
  {132002} (\bibinfo {year} {2015}{\natexlab{a}})},\ \Eprint
  {http://arxiv.org/abs/1507.03704} {arXiv:1507.03704 [hep-ph]} \BibitemShut
  {NoStop}%
\bibitem [{\citenamefont {Chen}\ \emph
  {et~al.}(2015{\natexlab{b}})\citenamefont {Chen}, \citenamefont {Chen},
  \citenamefont {Liu}, \citenamefont {Steele},\ and\ \citenamefont
  {Zhu}}]{Chen:2015moa}%
  \BibitemOpen
  \bibfield  {author} {\bibinfo {author} {\bibfnamefont {H.-X.}\ \bibnamefont
  {Chen}}, \bibinfo {author} {\bibfnamefont {W.}~\bibnamefont {Chen}}, \bibinfo
  {author} {\bibfnamefont {X.}~\bibnamefont {Liu}}, \bibinfo {author}
  {\bibfnamefont {T.~G.}\ \bibnamefont {Steele}}, \ and\ \bibinfo {author}
  {\bibfnamefont {S.-L.}\ \bibnamefont {Zhu}},\ }\href {\doibase
  10.1103/PhysRevLett.115.172001} {\bibfield  {journal} {\bibinfo  {journal}
  {Phys. Rev. Lett.}\ }\textbf {\bibinfo {volume} {115}},\ \bibinfo {pages}
  {172001} (\bibinfo {year} {2015}{\natexlab{b}})},\ \Eprint
  {http://arxiv.org/abs/1507.03717} {arXiv:1507.03717 [hep-ph]} \BibitemShut
  {NoStop}%
\bibitem [{\citenamefont {Roca}\ \emph {et~al.}(2015)\citenamefont {Roca},
  \citenamefont {Nieves},\ and\ \citenamefont {Oset}}]{Roca:2015dva}%
  \BibitemOpen
  \bibfield  {author} {\bibinfo {author} {\bibfnamefont {L.}~\bibnamefont
  {Roca}}, \bibinfo {author} {\bibfnamefont {J.}~\bibnamefont {Nieves}}, \ and\
  \bibinfo {author} {\bibfnamefont {E.}~\bibnamefont {Oset}},\ }\href {\doibase
  10.1103/PhysRevD.92.094003} {\bibfield  {journal} {\bibinfo  {journal} {Phys.
  Rev.}\ }\textbf {\bibinfo {volume} {D92}},\ \bibinfo {pages} {094003}
  (\bibinfo {year} {2015})},\ \Eprint {http://arxiv.org/abs/1507.04249}
  {arXiv:1507.04249 [hep-ph]} \BibitemShut {NoStop}%
\bibitem [{\citenamefont {He}(2016)}]{He:2015cea}%
  \BibitemOpen
  \bibfield  {author} {\bibinfo {author} {\bibfnamefont {J.}~\bibnamefont
  {He}},\ }\href {\doibase 10.1016/j.physletb.2015.12.071} {\bibfield
  {journal} {\bibinfo  {journal} {Phys. Lett.}\ }\textbf {\bibinfo {volume}
  {B753}},\ \bibinfo {pages} {547} (\bibinfo {year} {2016})},\ \Eprint
  {http://arxiv.org/abs/1507.05200} {arXiv:1507.05200 [hep-ph]} \BibitemShut
  {NoStop}%
\bibitem [{\citenamefont {Meißner}\ and\ \citenamefont
  {Oller}(2015)}]{Meissner:2015mza}%
  \BibitemOpen
  \bibfield  {author} {\bibinfo {author} {\bibfnamefont {U.-G.}\ \bibnamefont
  {Meißner}}\ and\ \bibinfo {author} {\bibfnamefont {J.~A.}\ \bibnamefont
  {Oller}},\ }\href {\doibase 10.1016/j.physletb.2015.10.015} {\bibfield
  {journal} {\bibinfo  {journal} {Phys. Lett.}\ }\textbf {\bibinfo {volume}
  {B751}},\ \bibinfo {pages} {59} (\bibinfo {year} {2015})},\ \Eprint
  {http://arxiv.org/abs/1507.07478} {arXiv:1507.07478 [hep-ph]} \BibitemShut
  {NoStop}%
\bibitem [{\citenamefont {Burns}(2015)}]{Burns:2015dwa}%
  \BibitemOpen
  \bibfield  {author} {\bibinfo {author} {\bibfnamefont {T.~J.}\ \bibnamefont
  {Burns}},\ }\href {\doibase 10.1140/epja/i2015-15152-6} {\bibfield  {journal}
  {\bibinfo  {journal} {Eur. Phys. J.}\ }\textbf {\bibinfo {volume} {A51}},\
  \bibinfo {pages} {152} (\bibinfo {year} {2015})},\ \Eprint
  {http://arxiv.org/abs/1509.02460} {arXiv:1509.02460 [hep-ph]} \BibitemShut
  {NoStop}%
\bibitem [{\citenamefont {Xiao}\ and\ \citenamefont
  {Meißner}(2015)}]{Xiao:2015fia}%
  \BibitemOpen
  \bibfield  {author} {\bibinfo {author} {\bibfnamefont {C.-W.}\ \bibnamefont
  {Xiao}}\ and\ \bibinfo {author} {\bibfnamefont {U.-G.}\ \bibnamefont
  {Meißner}},\ }\href {\doibase 10.1103/PhysRevD.92.114002} {\bibfield
  {journal} {\bibinfo  {journal} {Phys. Rev.}\ }\textbf {\bibinfo {volume}
  {D92}},\ \bibinfo {pages} {114002} (\bibinfo {year} {2015})},\ \Eprint
  {http://arxiv.org/abs/1508.00924} {arXiv:1508.00924 [hep-ph]} \BibitemShut
  {NoStop}%
\bibitem [{\citenamefont {Lü}\ and\ \citenamefont {Dong}(2016)}]{Lu:2016nnt}%
  \BibitemOpen
  \bibfield  {author} {\bibinfo {author} {\bibfnamefont {Q.-F.}\ \bibnamefont
  {Lü}}\ and\ \bibinfo {author} {\bibfnamefont {Y.-B.}\ \bibnamefont {Dong}},\
  }\href {\doibase 10.1103/PhysRevD.93.074020} {\bibfield  {journal} {\bibinfo
  {journal} {Phys. Rev.}\ }\textbf {\bibinfo {volume} {D93}},\ \bibinfo {pages}
  {074020} (\bibinfo {year} {2016})},\ \Eprint
  {http://arxiv.org/abs/1603.00559} {arXiv:1603.00559 [hep-ph]} \BibitemShut
  {NoStop}%
\bibitem [{\citenamefont {Shen}\ \emph {et~al.}(2016)\citenamefont {Shen},
  \citenamefont {Guo}, \citenamefont {Xie},\ and\ \citenamefont
  {Zou}}]{Shen:2016tzq}%
  \BibitemOpen
  \bibfield  {author} {\bibinfo {author} {\bibfnamefont {C.-W.}\ \bibnamefont
  {Shen}}, \bibinfo {author} {\bibfnamefont {F.-K.}\ \bibnamefont {Guo}},
  \bibinfo {author} {\bibfnamefont {J.-J.}\ \bibnamefont {Xie}}, \ and\
  \bibinfo {author} {\bibfnamefont {B.-S.}\ \bibnamefont {Zou}},\ }\href
  {\doibase 10.1016/j.nuclphysa.2016.04.034} {\bibfield  {journal} {\bibinfo
  {journal} {Nucl. Phys.}\ }\textbf {\bibinfo {volume} {A954}},\ \bibinfo
  {pages} {393} (\bibinfo {year} {2016})},\ \Eprint
  {http://arxiv.org/abs/1603.04672} {arXiv:1603.04672 [hep-ph]} \BibitemShut
  {NoStop}%
\bibitem [{\citenamefont {Yamaguchi}\ and\ \citenamefont
  {Santopinto}(2017)}]{Yamaguchi:2016ote}%
  \BibitemOpen
  \bibfield  {author} {\bibinfo {author} {\bibfnamefont {Y.}~\bibnamefont
  {Yamaguchi}}\ and\ \bibinfo {author} {\bibfnamefont {E.}~\bibnamefont
  {Santopinto}},\ }\href {\doibase 10.1103/PhysRevD.96.014018} {\bibfield
  {journal} {\bibinfo  {journal} {Phys. Rev.}\ }\textbf {\bibinfo {volume}
  {D96}},\ \bibinfo {pages} {014018} (\bibinfo {year} {2017})},\ \Eprint
  {http://arxiv.org/abs/1606.08330} {arXiv:1606.08330 [hep-ph]} \BibitemShut
  {NoStop}%
\bibitem [{\citenamefont {Azizi}\ \emph {et~al.}(2017)\citenamefont {Azizi},
  \citenamefont {Sarac},\ and\ \citenamefont {Sundu}}]{Azizi:2016dhy}%
  \BibitemOpen
  \bibfield  {author} {\bibinfo {author} {\bibfnamefont {K.}~\bibnamefont
  {Azizi}}, \bibinfo {author} {\bibfnamefont {Y.}~\bibnamefont {Sarac}}, \ and\
  \bibinfo {author} {\bibfnamefont {H.}~\bibnamefont {Sundu}},\ }\href
  {\doibase 10.1103/PhysRevD.95.094016} {\bibfield  {journal} {\bibinfo
  {journal} {Phys. Rev.}\ }\textbf {\bibinfo {volume} {D95}},\ \bibinfo {pages}
  {094016} (\bibinfo {year} {2017})},\ \Eprint
  {http://arxiv.org/abs/1612.07479} {arXiv:1612.07479 [hep-ph]} \BibitemShut
  {NoStop}%
\bibitem [{\citenamefont {Lin}\ \emph {et~al.}(2017)\citenamefont {Lin},
  \citenamefont {Shen}, \citenamefont {Guo},\ and\ \citenamefont
  {Zou}}]{Lin:2017mtz}%
  \BibitemOpen
  \bibfield  {author} {\bibinfo {author} {\bibfnamefont {Y.-H.}\ \bibnamefont
  {Lin}}, \bibinfo {author} {\bibfnamefont {C.-W.}\ \bibnamefont {Shen}},
  \bibinfo {author} {\bibfnamefont {F.-K.}\ \bibnamefont {Guo}}, \ and\
  \bibinfo {author} {\bibfnamefont {B.-S.}\ \bibnamefont {Zou}},\ }\href
  {\doibase 10.1103/PhysRevD.95.114017} {\bibfield  {journal} {\bibinfo
  {journal} {Phys. Rev.}\ }\textbf {\bibinfo {volume} {D95}},\ \bibinfo {pages}
  {114017} (\bibinfo {year} {2017})},\ \Eprint
  {http://arxiv.org/abs/1703.01045} {arXiv:1703.01045 [hep-ph]} \BibitemShut
  {NoStop}%
\bibitem [{\citenamefont {Geng}\ \emph {et~al.}(2018)\citenamefont {Geng},
  \citenamefont {Lu},\ and\ \citenamefont {Valderrama}}]{Geng:2017hxc}%
  \BibitemOpen
  \bibfield  {author} {\bibinfo {author} {\bibfnamefont {L.}~\bibnamefont
  {Geng}}, \bibinfo {author} {\bibfnamefont {J.}~\bibnamefont {Lu}}, \ and\
  \bibinfo {author} {\bibfnamefont {M.~P.}\ \bibnamefont {Valderrama}},\ }\href
  {\doibase 10.1103/PhysRevD.97.094036} {\bibfield  {journal} {\bibinfo
  {journal} {Phys. Rev.}\ }\textbf {\bibinfo {volume} {D97}},\ \bibinfo {pages}
  {094036} (\bibinfo {year} {2018})},\ \Eprint
  {http://arxiv.org/abs/1704.06123} {arXiv:1704.06123 [hep-ph]} \BibitemShut
  {NoStop}%
\bibitem [{\citenamefont {Yamaguchi}\ \emph {et~al.}(2017)\citenamefont
  {Yamaguchi}, \citenamefont {Giachino}, \citenamefont {Hosaka}, \citenamefont
  {Santopinto}, \citenamefont {Takeuchi},\ and\ \citenamefont
  {Takizawa}}]{Yamaguchi:2017zmn}%
  \BibitemOpen
  \bibfield  {author} {\bibinfo {author} {\bibfnamefont {Y.}~\bibnamefont
  {Yamaguchi}}, \bibinfo {author} {\bibfnamefont {A.}~\bibnamefont {Giachino}},
  \bibinfo {author} {\bibfnamefont {A.}~\bibnamefont {Hosaka}}, \bibinfo
  {author} {\bibfnamefont {E.}~\bibnamefont {Santopinto}}, \bibinfo {author}
  {\bibfnamefont {S.}~\bibnamefont {Takeuchi}}, \ and\ \bibinfo {author}
  {\bibfnamefont {M.}~\bibnamefont {Takizawa}},\ }\href {\doibase
  10.1103/PhysRevD.96.114031} {\bibfield  {journal} {\bibinfo  {journal} {Phys.
  Rev.}\ }\textbf {\bibinfo {volume} {D96}},\ \bibinfo {pages} {114031}
  (\bibinfo {year} {2017})},\ \Eprint {http://arxiv.org/abs/1709.00819}
  {arXiv:1709.00819 [hep-ph]} \BibitemShut {NoStop}%
\bibitem [{\citenamefont {Wang}(2018)}]{Wang:2018waa}%
  \BibitemOpen
  \bibfield  {author} {\bibinfo {author} {\bibfnamefont {Z.-G.}\ \bibnamefont
  {Wang}},\ }\href@noop {} {\  (\bibinfo {year} {2018})},\ \Eprint
  {http://arxiv.org/abs/1806.10384} {arXiv:1806.10384 [hep-ph]} \BibitemShut
  {NoStop}%
\bibitem [{\citenamefont {Liu}\ \emph {et~al.}(2018)\citenamefont {Liu},
  \citenamefont {Peng}, \citenamefont {Sánchez~Sánchez},\ and\ \citenamefont
  {Valderrama}}]{Liu:2018zzu}%
  \BibitemOpen
  \bibfield  {author} {\bibinfo {author} {\bibfnamefont {M.-Z.}\ \bibnamefont
  {Liu}}, \bibinfo {author} {\bibfnamefont {F.-Z.}\ \bibnamefont {Peng}},
  \bibinfo {author} {\bibfnamefont {M.}~\bibnamefont {Sánchez~Sánchez}}, \
  and\ \bibinfo {author} {\bibfnamefont {M.~P.}\ \bibnamefont {Valderrama}},\
  }\href {\doibase 10.1103/PhysRevD.98.114030} {\bibfield  {journal} {\bibinfo
  {journal} {Phys. Rev.}\ }\textbf {\bibinfo {volume} {D98}},\ \bibinfo {pages}
  {114030} (\bibinfo {year} {2018})},\ \Eprint
  {http://arxiv.org/abs/1811.03992} {arXiv:1811.03992 [hep-ph]} \BibitemShut
  {NoStop}%
\bibitem [{\citenamefont {Maiani}\ \emph {et~al.}(2015)\citenamefont {Maiani},
  \citenamefont {Polosa},\ and\ \citenamefont {Riquer}}]{Maiani:2015vwa}%
  \BibitemOpen
  \bibfield  {author} {\bibinfo {author} {\bibfnamefont {L.}~\bibnamefont
  {Maiani}}, \bibinfo {author} {\bibfnamefont {A.~D.}\ \bibnamefont {Polosa}},
  \ and\ \bibinfo {author} {\bibfnamefont {V.}~\bibnamefont {Riquer}},\ }\href
  {\doibase 10.1016/j.physletb.2015.08.008} {\bibfield  {journal} {\bibinfo
  {journal} {Phys. Lett.}\ }\textbf {\bibinfo {volume} {B749}},\ \bibinfo
  {pages} {289} (\bibinfo {year} {2015})},\ \Eprint
  {http://arxiv.org/abs/1507.04980} {arXiv:1507.04980 [hep-ph]} \BibitemShut
  {NoStop}%
\bibitem [{\citenamefont {Lebed}(2015)}]{Lebed:2015tna}%
  \BibitemOpen
  \bibfield  {author} {\bibinfo {author} {\bibfnamefont {R.~F.}\ \bibnamefont
  {Lebed}},\ }\href {\doibase 10.1016/j.physletb.2015.08.032} {\bibfield
  {journal} {\bibinfo  {journal} {Phys. Lett.}\ }\textbf {\bibinfo {volume}
  {B749}},\ \bibinfo {pages} {454} (\bibinfo {year} {2015})},\ \Eprint
  {http://arxiv.org/abs/1507.05867} {arXiv:1507.05867 [hep-ph]} \BibitemShut
  {NoStop}%
\bibitem [{\citenamefont {Li}\ \emph {et~al.}(2015)\citenamefont {Li},
  \citenamefont {He},\ and\ \citenamefont {He}}]{Li:2015gta}%
  \BibitemOpen
  \bibfield  {author} {\bibinfo {author} {\bibfnamefont {G.-N.}\ \bibnamefont
  {Li}}, \bibinfo {author} {\bibfnamefont {X.-G.}\ \bibnamefont {He}}, \ and\
  \bibinfo {author} {\bibfnamefont {M.}~\bibnamefont {He}},\ }\href {\doibase
  10.1007/JHEP12(2015)128} {\bibfield  {journal} {\bibinfo  {journal} {JHEP}\
  }\textbf {\bibinfo {volume} {12}},\ \bibinfo {pages} {128} (\bibinfo {year}
  {2015})},\ \Eprint {http://arxiv.org/abs/1507.08252} {arXiv:1507.08252
  [hep-ph]} \BibitemShut {NoStop}%
\bibitem [{\citenamefont {Mironov}\ and\ \citenamefont
  {Morozov}(2015)}]{Mironov:2015ica}%
  \BibitemOpen
  \bibfield  {author} {\bibinfo {author} {\bibfnamefont {A.}~\bibnamefont
  {Mironov}}\ and\ \bibinfo {author} {\bibfnamefont {A.}~\bibnamefont
  {Morozov}},\ }\href {\doibase 10.7868/S0370274X15170038,
  10.1134/S0021364015170099} {\bibfield  {journal} {\bibinfo  {journal} {JETP
  Lett.}\ }\textbf {\bibinfo {volume} {102}},\ \bibinfo {pages} {271} (\bibinfo
  {year} {2015})},\ \bibinfo {note} {[Pisma Zh. Eksp. Teor.
  Fiz.102,no.5,302(2015)]},\ \Eprint {http://arxiv.org/abs/1507.04694}
  {arXiv:1507.04694 [hep-ph]} \BibitemShut {NoStop}%
\bibitem [{\citenamefont {Anisovich}\ \emph {et~al.}(2015)\citenamefont
  {Anisovich}, \citenamefont {Matveev}, \citenamefont {Nyiri}, \citenamefont
  {Sarantsev},\ and\ \citenamefont {Semenova}}]{Anisovich:2015cia}%
  \BibitemOpen
  \bibfield  {author} {\bibinfo {author} {\bibfnamefont {V.~V.}\ \bibnamefont
  {Anisovich}}, \bibinfo {author} {\bibfnamefont {M.~A.}\ \bibnamefont
  {Matveev}}, \bibinfo {author} {\bibfnamefont {J.}~\bibnamefont {Nyiri}},
  \bibinfo {author} {\bibfnamefont {A.~V.}\ \bibnamefont {Sarantsev}}, \ and\
  \bibinfo {author} {\bibfnamefont {A.~N.}\ \bibnamefont {Semenova}},\
  }\href@noop {} {\  (\bibinfo {year} {2015})},\ \Eprint
  {http://arxiv.org/abs/1507.07652} {arXiv:1507.07652 [hep-ph]} \BibitemShut
  {NoStop}%
\bibitem [{\citenamefont {Wang}(2016)}]{Wang:2015epa}%
  \BibitemOpen
  \bibfield  {author} {\bibinfo {author} {\bibfnamefont {Z.-G.}\ \bibnamefont
  {Wang}},\ }\href {\doibase 10.1140/epjc/s10052-016-3920-4} {\bibfield
  {journal} {\bibinfo  {journal} {Eur. Phys. J.}\ }\textbf {\bibinfo {volume}
  {C76}},\ \bibinfo {pages} {70} (\bibinfo {year} {2016})},\ \Eprint
  {http://arxiv.org/abs/1508.01468} {arXiv:1508.01468 [hep-ph]} \BibitemShut
  {NoStop}%
\bibitem [{\citenamefont {Ali}\ \emph {et~al.}(2016)\citenamefont {Ali},
  \citenamefont {Ahmed}, \citenamefont {Aslam},\ and\ \citenamefont
  {Rehman}}]{Ali:2016dkf}%
  \BibitemOpen
  \bibfield  {author} {\bibinfo {author} {\bibfnamefont {A.}~\bibnamefont
  {Ali}}, \bibinfo {author} {\bibfnamefont {I.}~\bibnamefont {Ahmed}}, \bibinfo
  {author} {\bibfnamefont {M.~J.}\ \bibnamefont {Aslam}}, \ and\ \bibinfo
  {author} {\bibfnamefont {A.}~\bibnamefont {Rehman}},\ }\href {\doibase
  10.1103/PhysRevD.94.054001} {\bibfield  {journal} {\bibinfo  {journal} {Phys.
  Rev.}\ }\textbf {\bibinfo {volume} {D94}},\ \bibinfo {pages} {054001}
  (\bibinfo {year} {2016})},\ \Eprint {http://arxiv.org/abs/1607.00987}
  {arXiv:1607.00987 [hep-ph]} \BibitemShut {NoStop}%
\bibitem [{\citenamefont {Takeuchi}\ and\ \citenamefont
  {Takizawa}(2017)}]{Takeuchi:2016ejt}%
  \BibitemOpen
  \bibfield  {author} {\bibinfo {author} {\bibfnamefont {S.}~\bibnamefont
  {Takeuchi}}\ and\ \bibinfo {author} {\bibfnamefont {M.}~\bibnamefont
  {Takizawa}},\ }\href {\doibase 10.1016/j.physletb.2016.11.034} {\bibfield
  {journal} {\bibinfo  {journal} {Phys. Lett.}\ }\textbf {\bibinfo {volume}
  {B764}},\ \bibinfo {pages} {254} (\bibinfo {year} {2017})},\ \Eprint
  {http://arxiv.org/abs/1608.05475} {arXiv:1608.05475 [hep-ph]} \BibitemShut
  {NoStop}%
\bibitem [{\citenamefont {Kubarovsky}\ and\ \citenamefont
  {Voloshin}(2015)}]{Kubarovsky:2015aaa}%
  \BibitemOpen
  \bibfield  {author} {\bibinfo {author} {\bibfnamefont {V.}~\bibnamefont
  {Kubarovsky}}\ and\ \bibinfo {author} {\bibfnamefont {M.~B.}\ \bibnamefont
  {Voloshin}},\ }\href {\doibase 10.1103/PhysRevD.92.031502} {\bibfield
  {journal} {\bibinfo  {journal} {Phys. Rev.}\ }\textbf {\bibinfo {volume}
  {D92}},\ \bibinfo {pages} {031502} (\bibinfo {year} {2015})},\ \Eprint
  {http://arxiv.org/abs/1508.00888} {arXiv:1508.00888 [hep-ph]} \BibitemShut
  {NoStop}%
\bibitem [{\citenamefont {Guo}\ \emph {et~al.}(2015)\citenamefont {Guo},
  \citenamefont {Meißner}, \citenamefont {Wang},\ and\ \citenamefont
  {Yang}}]{Guo:2015umn}%
  \BibitemOpen
  \bibfield  {author} {\bibinfo {author} {\bibfnamefont {F.-K.}\ \bibnamefont
  {Guo}}, \bibinfo {author} {\bibfnamefont {U.-G.}\ \bibnamefont {Meißner}},
  \bibinfo {author} {\bibfnamefont {W.}~\bibnamefont {Wang}}, \ and\ \bibinfo
  {author} {\bibfnamefont {Z.}~\bibnamefont {Yang}},\ }\href {\doibase
  10.1103/PhysRevD.92.071502} {\bibfield  {journal} {\bibinfo  {journal} {Phys.
  Rev.}\ }\textbf {\bibinfo {volume} {D92}},\ \bibinfo {pages} {071502}
  (\bibinfo {year} {2015})},\ \Eprint {http://arxiv.org/abs/1507.04950}
  {arXiv:1507.04950 [hep-ph]} \BibitemShut {NoStop}%
\bibitem [{\citenamefont {Liu}\ \emph {et~al.}(2016)\citenamefont {Liu},
  \citenamefont {Wang},\ and\ \citenamefont {Zhao}}]{Liu:2015fea}%
  \BibitemOpen
  \bibfield  {author} {\bibinfo {author} {\bibfnamefont {X.-H.}\ \bibnamefont
  {Liu}}, \bibinfo {author} {\bibfnamefont {Q.}~\bibnamefont {Wang}}, \ and\
  \bibinfo {author} {\bibfnamefont {Q.}~\bibnamefont {Zhao}},\ }\href {\doibase
  10.1016/j.physletb.2016.03.089} {\bibfield  {journal} {\bibinfo  {journal}
  {Phys. Lett.}\ }\textbf {\bibinfo {volume} {B757}},\ \bibinfo {pages} {231}
  (\bibinfo {year} {2016})},\ \Eprint {http://arxiv.org/abs/1507.05359}
  {arXiv:1507.05359 [hep-ph]} \BibitemShut {NoStop}%
\bibitem [{\citenamefont {Guo}\ \emph {et~al.}(2016)\citenamefont {Guo},
  \citenamefont {Meißner}, \citenamefont {Nieves},\ and\ \citenamefont
  {Yang}}]{Guo:2016bkl}%
  \BibitemOpen
  \bibfield  {author} {\bibinfo {author} {\bibfnamefont {F.-K.}\ \bibnamefont
  {Guo}}, \bibinfo {author} {\bibfnamefont {U.-G.}\ \bibnamefont {Meißner}},
  \bibinfo {author} {\bibfnamefont {J.}~\bibnamefont {Nieves}}, \ and\ \bibinfo
  {author} {\bibfnamefont {Z.}~\bibnamefont {Yang}},\ }\href {\doibase
  10.1140/epja/i2016-16318-4} {\bibfield  {journal} {\bibinfo  {journal} {Eur.
  Phys. J.}\ }\textbf {\bibinfo {volume} {A52}},\ \bibinfo {pages} {318}
  (\bibinfo {year} {2016})},\ \Eprint {http://arxiv.org/abs/1605.05113}
  {arXiv:1605.05113 [hep-ph]} \BibitemShut {NoStop}%
\bibitem [{\citenamefont {Bayar}\ \emph {et~al.}(2016)\citenamefont {Bayar},
  \citenamefont {Aceti}, \citenamefont {Guo},\ and\ \citenamefont
  {Oset}}]{Bayar:2016ftu}%
  \BibitemOpen
  \bibfield  {author} {\bibinfo {author} {\bibfnamefont {M.}~\bibnamefont
  {Bayar}}, \bibinfo {author} {\bibfnamefont {F.}~\bibnamefont {Aceti}},
  \bibinfo {author} {\bibfnamefont {F.-K.}\ \bibnamefont {Guo}}, \ and\
  \bibinfo {author} {\bibfnamefont {E.}~\bibnamefont {Oset}},\ }\href {\doibase
  10.1103/PhysRevD.94.074039} {\bibfield  {journal} {\bibinfo  {journal} {Phys.
  Rev.}\ }\textbf {\bibinfo {volume} {D94}},\ \bibinfo {pages} {074039}
  (\bibinfo {year} {2016})},\ \Eprint {http://arxiv.org/abs/1609.04133}
  {arXiv:1609.04133 [hep-ph]} \BibitemShut {NoStop}%
\bibitem [{\citenamefont {Chen}\ \emph {et~al.}(2016)\citenamefont {Chen},
  \citenamefont {Chen}, \citenamefont {Liu},\ and\ \citenamefont
  {Zhu}}]{Chen:2016qju}%
  \BibitemOpen
  \bibfield  {author} {\bibinfo {author} {\bibfnamefont {H.-X.}\ \bibnamefont
  {Chen}}, \bibinfo {author} {\bibfnamefont {W.}~\bibnamefont {Chen}}, \bibinfo
  {author} {\bibfnamefont {X.}~\bibnamefont {Liu}}, \ and\ \bibinfo {author}
  {\bibfnamefont {S.-L.}\ \bibnamefont {Zhu}},\ }\href {\doibase
  10.1016/j.physrep.2016.05.004} {\bibfield  {journal} {\bibinfo  {journal}
  {Phys. Rept.}\ }\textbf {\bibinfo {volume} {639}},\ \bibinfo {pages} {1}
  (\bibinfo {year} {2016})},\ \Eprint {http://arxiv.org/abs/1601.02092}
  {arXiv:1601.02092 [hep-ph]} \BibitemShut {NoStop}%
\bibitem [{\citenamefont {Lebed}\ \emph {et~al.}(2017)\citenamefont {Lebed},
  \citenamefont {Mitchell},\ and\ \citenamefont {Swanson}}]{Lebed:2016hpi}%
  \BibitemOpen
  \bibfield  {author} {\bibinfo {author} {\bibfnamefont {R.~F.}\ \bibnamefont
  {Lebed}}, \bibinfo {author} {\bibfnamefont {R.~E.}\ \bibnamefont {Mitchell}},
  \ and\ \bibinfo {author} {\bibfnamefont {E.~S.}\ \bibnamefont {Swanson}},\
  }\href {\doibase 10.1016/j.ppnp.2016.11.003} {\bibfield  {journal} {\bibinfo
  {journal} {Prog. Part. Nucl. Phys.}\ }\textbf {\bibinfo {volume} {93}},\
  \bibinfo {pages} {143} (\bibinfo {year} {2017})},\ \Eprint
  {http://arxiv.org/abs/1610.04528} {arXiv:1610.04528 [hep-ph]} \BibitemShut
  {NoStop}%
\bibitem [{\citenamefont {Esposito}\ \emph {et~al.}(2016)\citenamefont
  {Esposito}, \citenamefont {Pilloni},\ and\ \citenamefont
  {Polosa}}]{Esposito:2016noz}%
  \BibitemOpen
  \bibfield  {author} {\bibinfo {author} {\bibfnamefont {A.}~\bibnamefont
  {Esposito}}, \bibinfo {author} {\bibfnamefont {A.}~\bibnamefont {Pilloni}}, \
  and\ \bibinfo {author} {\bibfnamefont {A.~D.}\ \bibnamefont {Polosa}},\
  }\href {\doibase 10.1016/j.physrep.2016.11.002} {\bibfield  {journal}
  {\bibinfo  {journal} {Phys. Rept.}\ }\textbf {\bibinfo {volume} {668}},\
  \bibinfo {pages} {1} (\bibinfo {year} {2016})},\ \Eprint
  {http://arxiv.org/abs/1611.07920} {arXiv:1611.07920 [hep-ph]} \BibitemShut
  {NoStop}%
\bibitem [{\citenamefont {Guo}\ \emph {et~al.}(2018{\natexlab{a}})\citenamefont
  {Guo}, \citenamefont {Hanhart}, \citenamefont {Meißner}, \citenamefont
  {Wang}, \citenamefont {Zhao},\ and\ \citenamefont {Zou}}]{Guo:2017jvc}%
  \BibitemOpen
  \bibfield  {author} {\bibinfo {author} {\bibfnamefont {F.-K.}\ \bibnamefont
  {Guo}}, \bibinfo {author} {\bibfnamefont {C.}~\bibnamefont {Hanhart}},
  \bibinfo {author} {\bibfnamefont {U.-G.}\ \bibnamefont {Meißner}}, \bibinfo
  {author} {\bibfnamefont {Q.}~\bibnamefont {Wang}}, \bibinfo {author}
  {\bibfnamefont {Q.}~\bibnamefont {Zhao}}, \ and\ \bibinfo {author}
  {\bibfnamefont {B.-S.}\ \bibnamefont {Zou}},\ }\href {\doibase
  10.1103/RevModPhys.90.015004} {\bibfield  {journal} {\bibinfo  {journal}
  {Rev. Mod. Phys.}\ }\textbf {\bibinfo {volume} {90}},\ \bibinfo {pages}
  {015004} (\bibinfo {year} {2018}{\natexlab{a}})},\ \Eprint
  {http://arxiv.org/abs/1705.00141} {arXiv:1705.00141 [hep-ph]} \BibitemShut
  {NoStop}%
\bibitem [{\citenamefont {Ali}\ \emph {et~al.}(2017)\citenamefont {Ali},
  \citenamefont {Lange},\ and\ \citenamefont {Stone}}]{Ali:2017jda}%
  \BibitemOpen
  \bibfield  {author} {\bibinfo {author} {\bibfnamefont {A.}~\bibnamefont
  {Ali}}, \bibinfo {author} {\bibfnamefont {J.~S.}\ \bibnamefont {Lange}}, \
  and\ \bibinfo {author} {\bibfnamefont {S.}~\bibnamefont {Stone}},\ }\href
  {\doibase 10.1016/j.ppnp.2017.08.003} {\bibfield  {journal} {\bibinfo
  {journal} {Prog. Part. Nucl. Phys.}\ }\textbf {\bibinfo {volume} {97}},\
  \bibinfo {pages} {123} (\bibinfo {year} {2017})},\ \Eprint
  {http://arxiv.org/abs/1706.00610} {arXiv:1706.00610 [hep-ph]} \BibitemShut
  {NoStop}%
\bibitem [{\citenamefont {Olsen}\ \emph {et~al.}(2018)\citenamefont {Olsen},
  \citenamefont {Skwarnicki},\ and\ \citenamefont {Zieminska}}]{Olsen:2017bmm}%
  \BibitemOpen
  \bibfield  {author} {\bibinfo {author} {\bibfnamefont {S.~L.}\ \bibnamefont
  {Olsen}}, \bibinfo {author} {\bibfnamefont {T.}~\bibnamefont {Skwarnicki}}, \
  and\ \bibinfo {author} {\bibfnamefont {D.}~\bibnamefont {Zieminska}},\ }\href
  {\doibase 10.1103/RevModPhys.90.015003} {\bibfield  {journal} {\bibinfo
  {journal} {Rev. Mod. Phys.}\ }\textbf {\bibinfo {volume} {90}},\ \bibinfo
  {pages} {015003} (\bibinfo {year} {2018})},\ \Eprint
  {http://arxiv.org/abs/1708.04012} {arXiv:1708.04012 [hep-ph]} \BibitemShut
  {NoStop}%
\bibitem [{\citenamefont {Karliner}\ \emph {et~al.}(2018)\citenamefont
  {Karliner}, \citenamefont {Rosner},\ and\ \citenamefont
  {Skwarnicki}}]{Karliner:2017qhf}%
  \BibitemOpen
  \bibfield  {author} {\bibinfo {author} {\bibfnamefont {M.}~\bibnamefont
  {Karliner}}, \bibinfo {author} {\bibfnamefont {J.~L.}\ \bibnamefont
  {Rosner}}, \ and\ \bibinfo {author} {\bibfnamefont {T.}~\bibnamefont
  {Skwarnicki}},\ }\href {\doibase 10.1146/annurev-nucl-101917-020902}
  {\bibfield  {journal} {\bibinfo  {journal} {Ann. Rev. Nucl. Part. Sci.}\
  }\textbf {\bibinfo {volume} {68}},\ \bibinfo {pages} {17} (\bibinfo {year}
  {2018})},\ \Eprint {http://arxiv.org/abs/1711.10626} {arXiv:1711.10626
  [hep-ph]} \BibitemShut {NoStop}%
\bibitem [{\citenamefont {Cerri}\ \emph {et~al.}(2018)\citenamefont {Cerri}
  \emph {et~al.}}]{Cerri:2018ypt}%
  \BibitemOpen
  \bibfield  {author} {\bibinfo {author} {\bibfnamefont {A.}~\bibnamefont
  {Cerri}} \emph {et~al.},\ }\href@noop {} {\  (\bibinfo {year} {2018})},\
  \Eprint {http://arxiv.org/abs/1812.07638} {arXiv:1812.07638 [hep-ph]}
  \BibitemShut {NoStop}%
\bibitem [{\citenamefont {Chen}\ \emph
  {et~al.}(2019{\natexlab{a}})\citenamefont {Chen}, \citenamefont {Chen},\ and\
  \citenamefont {Zhu}}]{Chen:2019bip}%
  \BibitemOpen
  \bibfield  {author} {\bibinfo {author} {\bibfnamefont {H.-X.}\ \bibnamefont
  {Chen}}, \bibinfo {author} {\bibfnamefont {W.}~\bibnamefont {Chen}}, \ and\
  \bibinfo {author} {\bibfnamefont {S.-L.}\ \bibnamefont {Zhu}},\ }\href@noop
  {} {\  (\bibinfo {year} {2019}{\natexlab{a}})},\ \Eprint
  {http://arxiv.org/abs/1903.11001} {arXiv:1903.11001 [hep-ph]} \BibitemShut
  {NoStop}%
\bibitem [{\citenamefont {Chen}\ \emph
  {et~al.}(2019{\natexlab{b}})\citenamefont {Chen}, \citenamefont {Liu},
  \citenamefont {Sun},\ and\ \citenamefont {Zhu}}]{Chen:2019asm}%
  \BibitemOpen
  \bibfield  {author} {\bibinfo {author} {\bibfnamefont {R.}~\bibnamefont
  {Chen}}, \bibinfo {author} {\bibfnamefont {X.}~\bibnamefont {Liu}}, \bibinfo
  {author} {\bibfnamefont {Z.-F.}\ \bibnamefont {Sun}}, \ and\ \bibinfo
  {author} {\bibfnamefont {S.-L.}\ \bibnamefont {Zhu}},\ }\href@noop {} {\
  (\bibinfo {year} {2019}{\natexlab{b}})},\ \Eprint
  {http://arxiv.org/abs/1903.11013} {arXiv:1903.11013 [hep-ph]} \BibitemShut
  {NoStop}%
\bibitem [{\citenamefont {Gamermann}\ \emph {et~al.}(2010)\citenamefont
  {Gamermann}, \citenamefont {Nieves}, \citenamefont {Oset},\ and\
  \citenamefont {Ruiz~Arriola}}]{Gamermann:2009uq}%
  \BibitemOpen
  \bibfield  {author} {\bibinfo {author} {\bibfnamefont {D.}~\bibnamefont
  {Gamermann}}, \bibinfo {author} {\bibfnamefont {J.}~\bibnamefont {Nieves}},
  \bibinfo {author} {\bibfnamefont {E.}~\bibnamefont {Oset}}, \ and\ \bibinfo
  {author} {\bibfnamefont {E.}~\bibnamefont {Ruiz~Arriola}},\ }\href {\doibase
  10.1103/PhysRevD.81.014029} {\bibfield  {journal} {\bibinfo  {journal} {Phys.
  Rev.}\ }\textbf {\bibinfo {volume} {D81}},\ \bibinfo {pages} {014029}
  (\bibinfo {year} {2010})},\ \Eprint {http://arxiv.org/abs/0911.4407}
  {arXiv:0911.4407 [hep-ph]} \BibitemShut {NoStop}%
\bibitem [{\citenamefont {Hanhart}\ \emph {et~al.}(2012)\citenamefont
  {Hanhart}, \citenamefont {Kalashnikova}, \citenamefont {Kudryavtsev},\ and\
  \citenamefont {Nefediev}}]{Hanhart:2011tn}%
  \BibitemOpen
  \bibfield  {author} {\bibinfo {author} {\bibfnamefont {C.}~\bibnamefont
  {Hanhart}}, \bibinfo {author} {\bibfnamefont {{\relax Yu}.~S.}\ \bibnamefont
  {Kalashnikova}}, \bibinfo {author} {\bibfnamefont {A.~E.}\ \bibnamefont
  {Kudryavtsev}}, \ and\ \bibinfo {author} {\bibfnamefont {A.~V.}\ \bibnamefont
  {Nefediev}},\ }\href {\doibase 10.1103/PhysRevD.85.011501} {\bibfield
  {journal} {\bibinfo  {journal} {Phys. Rev.}\ }\textbf {\bibinfo {volume}
  {D85}},\ \bibinfo {pages} {011501} (\bibinfo {year} {2012})},\ \Eprint
  {http://arxiv.org/abs/1111.6241} {arXiv:1111.6241 [hep-ph]} \BibitemShut
  {NoStop}%
\bibitem [{\citenamefont {Li}\ and\ \citenamefont {Zhu}(2012)}]{Li:2012cs}%
  \BibitemOpen
  \bibfield  {author} {\bibinfo {author} {\bibfnamefont {N.}~\bibnamefont
  {Li}}\ and\ \bibinfo {author} {\bibfnamefont {S.-L.}\ \bibnamefont {Zhu}},\
  }\href {\doibase 10.1103/PhysRevD.86.074022} {\bibfield  {journal} {\bibinfo
  {journal} {Phys. Rev.}\ }\textbf {\bibinfo {volume} {D86}},\ \bibinfo {pages}
  {074022} (\bibinfo {year} {2012})},\ \Eprint {http://arxiv.org/abs/1207.3954}
  {arXiv:1207.3954 [hep-ph]} \BibitemShut {NoStop}%
\bibitem [{\citenamefont {Dashen}\ \emph {et~al.}(1994)\citenamefont {Dashen},
  \citenamefont {Jenkins},\ and\ \citenamefont {Manohar}}]{Dashen:1993jt}%
  \BibitemOpen
  \bibfield  {author} {\bibinfo {author} {\bibfnamefont {R.~F.}\ \bibnamefont
  {Dashen}}, \bibinfo {author} {\bibfnamefont {E.~E.}\ \bibnamefont {Jenkins}},
  \ and\ \bibinfo {author} {\bibfnamefont {A.~V.}\ \bibnamefont {Manohar}},\
  }\href {\doibase 10.1103/PhysRevD.51.2489, 10.1103/PhysRevD.49.4713}
  {\bibfield  {journal} {\bibinfo  {journal} {Phys. Rev.}\ }\textbf {\bibinfo
  {volume} {D49}},\ \bibinfo {pages} {4713} (\bibinfo {year} {1994})},\
  \bibinfo {note} {[Erratum: Phys. Rev.D51,2489(1995)]},\ \Eprint
  {http://arxiv.org/abs/hep-ph/9310379} {arXiv:hep-ph/9310379 [hep-ph]}
  \BibitemShut {NoStop}%
\bibitem [{\citenamefont {Guo}\ \emph {et~al.}(2014)\citenamefont {Guo},
  \citenamefont {Hidalgo-Duque}, \citenamefont {Nieves}, \citenamefont
  {Ozpineci},\ and\ \citenamefont {Valderrama}}]{Guo:2014hqa}%
  \BibitemOpen
  \bibfield  {author} {\bibinfo {author} {\bibfnamefont {F.~K.}\ \bibnamefont
  {Guo}}, \bibinfo {author} {\bibfnamefont {C.}~\bibnamefont {Hidalgo-Duque}},
  \bibinfo {author} {\bibfnamefont {J.}~\bibnamefont {Nieves}}, \bibinfo
  {author} {\bibfnamefont {A.}~\bibnamefont {Ozpineci}}, \ and\ \bibinfo
  {author} {\bibfnamefont {M.~P.}\ \bibnamefont {Valderrama}},\ }\href
  {\doibase 10.1140/epjc/s10052-014-2885-4} {\bibfield  {journal} {\bibinfo
  {journal} {Eur. Phys. J.}\ }\textbf {\bibinfo {volume} {C74}},\ \bibinfo
  {pages} {2885} (\bibinfo {year} {2014})},\ \Eprint
  {http://arxiv.org/abs/1404.1776} {arXiv:1404.1776 [hep-ph]} \BibitemShut
  {NoStop}%
\bibitem [{\citenamefont {Faessler}\ \emph {et~al.}(2007)\citenamefont
  {Faessler}, \citenamefont {Gutsche}, \citenamefont {Lyubovitskij},\ and\
  \citenamefont {Ma}}]{Faessler:2007gv}%
  \BibitemOpen
  \bibfield  {author} {\bibinfo {author} {\bibfnamefont {A.}~\bibnamefont
  {Faessler}}, \bibinfo {author} {\bibfnamefont {T.}~\bibnamefont {Gutsche}},
  \bibinfo {author} {\bibfnamefont {V.~E.}\ \bibnamefont {Lyubovitskij}}, \
  and\ \bibinfo {author} {\bibfnamefont {Y.-L.}\ \bibnamefont {Ma}},\ }\href
  {\doibase 10.1103/PhysRevD.76.014005} {\bibfield  {journal} {\bibinfo
  {journal} {Phys. Rev.}\ }\textbf {\bibinfo {volume} {D76}},\ \bibinfo {pages}
  {014005} (\bibinfo {year} {2007})},\ \Eprint {http://arxiv.org/abs/0705.0254}
  {arXiv:0705.0254 [hep-ph]} \BibitemShut {NoStop}%
\bibitem [{\citenamefont {Lutz}\ and\ \citenamefont
  {Soyeur}(2008)}]{Lutz:2007sk}%
  \BibitemOpen
  \bibfield  {author} {\bibinfo {author} {\bibfnamefont {M.~F.~M.}\
  \bibnamefont {Lutz}}\ and\ \bibinfo {author} {\bibfnamefont {M.}~\bibnamefont
  {Soyeur}},\ }\href {\doibase 10.1016/j.nuclphysa.2008.09.003} {\bibfield
  {journal} {\bibinfo  {journal} {Nucl. Phys.}\ }\textbf {\bibinfo {volume}
  {A813}},\ \bibinfo {pages} {14} (\bibinfo {year} {2008})},\ \Eprint
  {http://arxiv.org/abs/0710.1545} {arXiv:0710.1545 [hep-ph]} \BibitemShut
  {NoStop}%
\bibitem [{\citenamefont {Guo}\ \emph {et~al.}(2008)\citenamefont {Guo},
  \citenamefont {Hanhart}, \citenamefont {Krewald},\ and\ \citenamefont
  {Mei{\ss}ner}}]{Guo:2008gp}%
  \BibitemOpen
  \bibfield  {author} {\bibinfo {author} {\bibfnamefont {F.-K.}\ \bibnamefont
  {Guo}}, \bibinfo {author} {\bibfnamefont {C.}~\bibnamefont {Hanhart}},
  \bibinfo {author} {\bibfnamefont {S.}~\bibnamefont {Krewald}}, \ and\
  \bibinfo {author} {\bibfnamefont {U.-G.}\ \bibnamefont {Mei{\ss}ner}},\
  }\href {\doibase 10.1016/j.physletb.2008.07.060} {\bibfield  {journal}
  {\bibinfo  {journal} {Phys. Lett.}\ }\textbf {\bibinfo {volume} {B666}},\
  \bibinfo {pages} {251} (\bibinfo {year} {2008})},\ \Eprint
  {http://arxiv.org/abs/0806.3374} {arXiv:0806.3374 [hep-ph]} \BibitemShut
  {NoStop}%
\bibitem [{\citenamefont {Liu}\ \emph {et~al.}(2013)\citenamefont {Liu},
  \citenamefont {Orginos}, \citenamefont {Guo}, \citenamefont {Hanhart},\ and\
  \citenamefont {Mei{\ss}ner}}]{Liu:2012zya}%
  \BibitemOpen
  \bibfield  {author} {\bibinfo {author} {\bibfnamefont {L.}~\bibnamefont
  {Liu}}, \bibinfo {author} {\bibfnamefont {K.}~\bibnamefont {Orginos}},
  \bibinfo {author} {\bibfnamefont {F.-K.}\ \bibnamefont {Guo}}, \bibinfo
  {author} {\bibfnamefont {C.}~\bibnamefont {Hanhart}}, \ and\ \bibinfo
  {author} {\bibfnamefont {U.-G.}\ \bibnamefont {Mei{\ss}ner}},\ }\href
  {\doibase 10.1103/PhysRevD.87.014508} {\bibfield  {journal} {\bibinfo
  {journal} {Phys. Rev.}\ }\textbf {\bibinfo {volume} {D87}},\ \bibinfo {pages}
  {014508} (\bibinfo {year} {2013})},\ \Eprint {http://arxiv.org/abs/1208.4535}
  {arXiv:1208.4535 [hep-lat]} \BibitemShut {NoStop}%
\bibitem [{\citenamefont {Guo}\ \emph {et~al.}(2018{\natexlab{b}})\citenamefont
  {Guo}, \citenamefont {Heo},\ and\ \citenamefont {Lutz}}]{Guo:2018kno}%
  \BibitemOpen
  \bibfield  {author} {\bibinfo {author} {\bibfnamefont {X.-Y.}\ \bibnamefont
  {Guo}}, \bibinfo {author} {\bibfnamefont {Y.}~\bibnamefont {Heo}}, \ and\
  \bibinfo {author} {\bibfnamefont {M.~F.~M.}\ \bibnamefont {Lutz}},\ }\href
  {\doibase 10.1103/PhysRevD.98.014510} {\bibfield  {journal} {\bibinfo
  {journal} {Phys. Rev.}\ }\textbf {\bibinfo {volume} {D98}},\ \bibinfo {pages}
  {014510} (\bibinfo {year} {2018}{\natexlab{b}})},\ \Eprint
  {http://arxiv.org/abs/1801.10122} {arXiv:1801.10122 [hep-lat]} \BibitemShut
  {NoStop}%
\bibitem [{\citenamefont {Bardeen}\ \emph {et~al.}(2003)\citenamefont
  {Bardeen}, \citenamefont {Eichten},\ and\ \citenamefont
  {Hill}}]{Bardeen:2003kt}%
  \BibitemOpen
  \bibfield  {author} {\bibinfo {author} {\bibfnamefont {W.~A.}\ \bibnamefont
  {Bardeen}}, \bibinfo {author} {\bibfnamefont {E.~J.}\ \bibnamefont
  {Eichten}}, \ and\ \bibinfo {author} {\bibfnamefont {C.~T.}\ \bibnamefont
  {Hill}},\ }\href {\doibase 10.1103/PhysRevD.68.054024} {\bibfield  {journal}
  {\bibinfo  {journal} {Phys. Rev.}\ }\textbf {\bibinfo {volume} {D68}},\
  \bibinfo {pages} {054024} (\bibinfo {year} {2003})},\ \Eprint
  {http://arxiv.org/abs/hep-ph/0305049} {arXiv:hep-ph/0305049 [hep-ph]}
  \BibitemShut {NoStop}%
\bibitem [{\citenamefont {Colangelo}\ and\ \citenamefont
  {De~Fazio}(2003)}]{Colangelo:2003vg}%
  \BibitemOpen
  \bibfield  {author} {\bibinfo {author} {\bibfnamefont {P.}~\bibnamefont
  {Colangelo}}\ and\ \bibinfo {author} {\bibfnamefont {F.}~\bibnamefont
  {De~Fazio}},\ }\href {\doibase 10.1016/j.physletb.2003.08.003} {\bibfield
  {journal} {\bibinfo  {journal} {Phys. Lett.}\ }\textbf {\bibinfo {volume}
  {B570}},\ \bibinfo {pages} {180} (\bibinfo {year} {2003})},\ \Eprint
  {http://arxiv.org/abs/hep-ph/0305140} {arXiv:hep-ph/0305140 [hep-ph]}
  \BibitemShut {NoStop}%
\end{thebibliography}%

\end{document}